\documentclass{INTERSPEECH2023}

\usepackage{physics}
\usepackage{amssymb}
\usepackage{makecell, graphics}
\usepackage{algorithm}
\usepackage{algorithmic}
\usepackage[misc]{ifsym} 

\interspeechcameraready

\title{Variance-Preserving-Based Interpolation Diffusion Models \\ for Speech Enhancement }
\name{Zilu Guo$^1$, Jun Du$^{1, *}$, Chin-Hui Lee$^2$, Yu Gao$^{3}$, Wenbin Zhang$^{3}$ \thanks{ * corresponding author}}
\address{
  $^1$University of Science and Technology of China, Hefei 230027, China\\
  $^2$Georgia Institute of Technology, Atlanta, GA. 30332-0250, USA\\
  $^3$AI Innovation Center, Midea Group (Shanghai) Co.,Ltd., Shanghai 201702, China
}
\email{guozl@mail.ustc.edu.cn, 
\textrm{\Letter}jundu@ustc.edu.cn, chl@ece.gatech.edu, \\  \{gaoyu11, zhangwb87\}@midea.com
}

\begin{document}

\maketitle

\begin{abstract}
The goal of this study is to implement diffusion models for speech enhancement (SE). The first step is to emphasize the theoretical foundation of variance-preserving (VP)-based interpolation diffusion under continuous conditions. Subsequently, we present a more concise framework that encapsulates both the VP- and variance-exploding (VE)-based interpolation diffusion methods. We demonstrate that these two methods are special cases of the proposed framework. Additionally, we provide a practical example of VP-based interpolation diffusion for the SE task. To improve performance and ease model training, we analyze the common difficulties encountered in diffusion models and suggest amenable hyper-parameters. Finally, we evaluate our model against several methods using a public benchmark to showcase the effectiveness of our approach. 

\end{abstract}

\section{Introduction}


Speech enhancement (SE) \cite{boll1979suppression}  has  been the subject of research for several decades with the goal  of diminishing or even  eliminating the noise in a noisy speech  while minimizing the distortion to speech quality. In recent years, SE has been approached as a supervised regressive task with the assistance of deep learning. 
The early attempts to deploy deep learning for SE involve employing  off-the-shelf models to predict the targets that are typically utilized in traditional approaches. 
However, it is important to note that these approaches are sub-optimal since they are not able to obtain the clean phase of the speech. 
 Common targets for these methods include the magnitude of the spectrogram \cite{wang2014training},  the log  magnitude (Mapping) \cite{xu2014regression}, the ideal ratio mask (IRM)\cite{narayanan2013ideal}, the spectral magnitude mask (SMM) \cite{wang2014training}, 
and etc.
The real and imaginary parts of the spectrogram are utilized directly as the target to obtain the clean phase afterward \cite{williamson_complex_2016, tan_complex_2019, choi2018phase}. Meanwhile, another main-stream SE method seeks to predict clean waveforms in an end-to-end (E2E) manner \cite{8331910}, rather than the spectrum. In addition to the regressive methods, some researchers have utilized  generative models for the SE, such as, VAE \cite{fang2021variational, Bando2020}, GAN \cite{Pascual2017, fu2019metricgan, fu21_interspeech}, flow \cite{nugraha2020flow, strauss2021flow}, and etc.

Recently, Diffusion models have been successful in various generative tasks, including image generation \cite{ho2020denoising, song2020score, song2019generative}, image editing \cite{kawar2023imagic}, speech synthesis \cite{diff-wav} and etc. Diffusion models involve two processes, i.e., diffusion (or forward)  and reverse (or backward) processes. Another approach in the discrete domain is the score-based model  \cite{song2019generative}.  Both the diffusion \cite{ho2020denoising}  and the score-based \cite{song2019generative, kingma2010} model  are unified in \cite{song2020score, kingma2021variational}, and generalized  to the continuous time domain, endowing the model with more capacity. Moreover, the authors in \cite{song2020score} have classified the diffusion 
 models into two types:  the  variance-preserving(VP)-based \cite{ho2020denoising}  and the variance-exploding(VE)-based \cite{song2019generative} method grounded on their intuitive properties.   The VE-based approach gradually increases the variance 
over time while keeping the clean component unaltered. In contrast, VP-based diffusion attempts to preserve the amplitude 
with fewer fluctuations. 
Besides their applications in the image processing field,  diffusion models have also been explored  for SE tasks.

In their work, CDiffSE  \cite{lu2022conditional} proposes that a degraded signal 
is composed of three components, i.e, the clean signal 
, the noisy 
and the 
Gaussian noise.  They suggest that the  mean of
the degraded signal in the diffusion process is obtained through a linear combination of the clean and noisy speech, specifically the linear interpolation of the two. They then apply this interpolation method to  Diffwave \cite{diff-wav},  a model for speech synthesis, for the SE task.
SRTNET \cite{srtnet} follows a similar approach, utilizing the interpolation diffusion to estimate the distortion between the clean speech and an enhanced speech predicted by a pre-trained discriminative model. However,  the two-stage models commonly have higher computational overhead.
Apart from the discrete diffusion models mentioned above, there are also several interpolation-based methods for SE under continuous conditions. In \cite{welker2022speech,richter2022speech}, the authors formulate the theoretical foundation for the  VE-based interpolation diffusion model (VEIDM) in the continuous time field and generalize it to the continuous time system. However, they leave the VP-based one untouched which showcases better performance in some tasks than the VE-based diffusion. In this paper, we attempt to apply linear interpolation to  VP-based diffusion.


The rest of the paper is organized as follows. The proposed method is introduced in Sec. \ref{proposed}. Specifically, we accentuate the proposed signal model, training method, and sampling algorithms in this section. Experiments settings, results, and analyses are presented in Sec. \ref{exp}. we draw conclusions in Sec. \ref{conclude}.

\section{The proposed method}
\label{proposed}
In the forward process of the vanilla VP-based diffusion model, a clean signal (such as an image or speech)  is gradually degraded by adding Gaussian noise in steps until it reaches an approximate Gaussian noise level. In this process,  the mean of the state $t$ is an affine function of the clean signal. However, in the VP-based interpolation diffusion, the mean is replaced with a linear interpolation of the clean and the noisy.

\begin{algorithm}
\caption{Training stage}
\label{training}
    \begin{algorithmic}
    \STATE {Batch size is $B$, number of training iterations $N_{\text{iter}}$}
    \FOR{$i=1$ to $N_{\text{iter}}$}
        \STATE {Sample a batch of  clean and noisy speech pairs [$(\pmb{x}_0^1, \pmb{y}^1)$, \dots, $(\pmb{x}_0^b, \pmb{y}^b)$, ..., $(\pmb{x}_0^B, \pmb{y}^B)]$,   $(\pmb{x}_0^b, \pmb{y}^b))\sim p(\pmb{x_0}, \pmb{y})$}
        \STATE {Sample a batch of the time indexes [$t^1$, $t^2$, ..., $t^B]$,   $t^b \sim \mathcal{U}(\epsilon, 1)$}
        \STATE {Sample a batch of Gaussian noises [$\pmb{z}^1$, $\pmb{z}^2$, ..., $\pmb{z}^B]$,   $\pmb{z}^b \sim \pmb{\mathcal{N}}(\pmb{0}, \pmb{I})$}
        \STATE {Get a batch of [$\pmb{x}^1(t^1)$, $\pmb{x}^2(t^2)$, ..., $\pmb{x}^B(t^2)$],    $\pmb{x}^b(t^b)$ is sampled from Eq.\eqref{equation:10}}
        \STATE {Input the 
        [$\pmb{x}^1(t^1)$, $\pmb{x}^2(t^2)$, ..., $\pmb{x}^B(t^2)$]
        to the neural network}
        \STATE {Get the  output [$\pmb{\theta}(\pmb{x}^1(t^1))$, $\pmb{\theta}(\pmb{x}^2(t^2))$,\dots, $\pmb{\theta}(\pmb{x}^B(t^B))$]}
        \STATE {Compute the loss $\mathcal{L} = \{\Sigma_{b=1}^B{||G(t^b)\pmb{\theta}(\pmb{x}(t^b)+{\pmb{z}}||^2}\}/B$ from Eq.\eqref{loss}}
        \STATE{Update the parameters}
    \ENDFOR
    
\end{algorithmic}
\end{algorithm}

\subsection{The VP-based interpolation diffusion model (VPIDM)}

For tasks of SE, image editing, voice conversion and etc, there is an existing condition that holds copious information about the target. In the case of SE, for instance,  the off-the-shelf noisy speech could be implemented  to guide the diffusion process of the clean.   
We refer to this approach as the VP-based interpolation diffusion model (VPIDM).   The signal model for VPIDM is defined as
\begin{align}
  \pmb{x}(t) = {\alpha}_t[\lambda _t\pmb{x}_0+(1-\lambda _t)\pmb{y}] + \sqrt{1 - {\alpha}_t^2}\pmb{z}
  \label{equation:10}
\end{align}
where $x(t)$ is the degraded signal at $t$ time index in the diffusion process, $\pmb{x}_0$ is the clean signal, $\pmb{y}$ is the noisy speech, $\pmb{z}$ is the Gaussian noise sampled from normal distribution, $\alpha_t$ determines the diffusion process, $\lambda_t$ is the slope of the interpolation process, $\alpha_t$ and $\lambda_t$ are functions of $t$. The differential of $\pmb{x}(t)$ is 
\begin{align}
  d\pmb{x}(t) = [ \pmb{x}(t)\ln^{'}{(\alpha}_t\lambda _t)-\pmb{y}{\alpha}_t\ln^{'}\lambda _t]dt + \pmb{\Sigma} _td\pmb{w}
  \label{equation:11}
\end{align}
where $\pmb{w}$ is the stochastic process, $ln^{'}[\cdot]$ is the derivative of $ln[\cdot]$ with respect to $t$, $d[\cdot]$ is the operation of differential. When $\pmb{\Sigma} _t$ is a  diagonal matrix, suppose that $\pmb{\Sigma} _t=g(t)\pmb{I}$. From Eq.(5.53) in \cite{solin_2019}, we get 
\begin{align}
    \frac{dG^2(t)}{dt} = 2G^2(t)\ln^{'}({\alpha}_t\lambda _t) + g^2(t)
    \label{equation:12}
\end{align}
where $g(t)$ indicates the spread speed of the stochastic process in the derivative of $\pmb{x}(t)$, $g(t) = \sqrt{-2 G^2(t)\ln^{'}\lambda_t - 2\ln^{'} \alpha_t}$, $G(t)$ is the coefficient of the Gaussian noise in $\pmb{x}(t)$, $G^2(t)=1-\alpha^2_t$, $t\in (0,1]$, $\ln^{'} \alpha_t\leq0$ and $\ln^{'}\lambda_t\leq0$, which means $\lambda_t$ and $\alpha_t$ are monotonous decrease functions. In this article, all constant superscripts  represent powers, unless otherwise specified. When $t\rightarrow 0$, then $\lambda_t\rightarrow 1$,$\alpha_t\rightarrow1$. 

In principle, we hope $\lambda_1 \rightarrow 0$, which implies that the final state is a combination of the noisy signal and the Gaussian noise. Therefore,  the larger $-\ln^{'}\lambda_t$ appears to be more favorable. However, empirical evidence suggests that $-\ln^{'}\lambda_t$ cannot be infinitely large.  The reason behind this is that when we sample a clean and  $-\ln^{'}\lambda_t$ is set to sufficiently large, the linear interpolation tends to change quickly  from   $\pmb{y}$ to $\pmb{x}_0$ over several steps, it is difficult for neural networks to capture.

\begin{algorithm}
\caption{Sampling (enhancing) stage}
\label{sampling}
    \begin{algorithmic}
    \STATE {Suppose the number of sample steps is $K$, $t_k = \frac{(1-\epsilon)}{K}k + \epsilon$}
    \STATE{Sample $\pmb{x}_K =\pmb{x}(t_K)= \pmb{x}(1)= \alpha_1\pmb{y}+G(1)\pmb{z}$}
    \FOR{$k=K-1, K-2, \dots, 1$}
    \STATE {Input the $\pmb{x}_{k+1}$, get the $\hat{\pmb{\theta}}(\pmb{x}_{k+1})$ }
    \STATE{ Sample a Gaussian noise $\pmb{z}$, $\pmb{z} \sim \pmb{\mathcal{N}}(\pmb{0}, \pmb{I})$}
    \STATE {Compute the $\pmb{x}_{k}$ from Eq.\eqref{resample}}
    
    \ENDFOR
\STATE {Input the $\pmb{x}_{1}$, get the $\hat{\pmb{\theta}}(\pmb{x}_{1})$}
    \STATE{$\hat{\pmb{x}}_{0} =  \pmb{x}_{1} - [f(\pmb{x}_1, \pmb{y}) -g_1^2\hat{\pmb{\theta}}(\pmb{x}_1) ]\Delta$}

\RETURN {$\hat{\pmb{x}}_{0}$}
\end{algorithmic}
\end{algorithm}

In this paper, we adopt the similar $\alpha_t$ schedule of the VP-based diffusion in \cite{song2020score} for the SE, i.e., $\alpha_t=e^{-0.5\int_0^t\beta(\tau)d\tau}$, where $\beta(t) =(\beta_{\text{max}}-\beta_{\text{min}})t+\beta_{\text{min}}$, $\beta_{\text{min}}$ controls the slope of the clean scale when $t\rightarrow0$, $(\beta_{\text{max}}-\beta_{\text{min}})$ controls the changing speed of $\pmb{x}_t$ from the clean to the Gaussian.
\begin{align}
    g(t) = \sqrt{\beta(t)+2\lambda(1-e^{-\int_0^t\beta(\tau)d\tau})}
\end{align}
where $\lambda_t=e^{-\lambda t}$. From Eq.\eqref{equation:11}, the $d\pmb{x}(t)$
\begin{align}
    d\pmb{x}(t) = [-(0.5\beta(t)+\lambda)\pmb{x}(t) + \lambda \alpha_t \pmb{y}]dt + g(t)d\pmb{w}
\end{align}
In the reverse process, the $\pmb{x}(1)$ is sampled from the distribution $\mathcal{N}(\alpha_1\pmb{y}, \sqrt{1-\alpha_1^2}\pmb{I})$.

%

\subsection{The loss function and the training stage}
For unconditional diffusion, the neural network is trained for predicting $\pmb{\nabla}_{\pmb{x}}\ln(p_{t}(\pmb{x}))$. This is equivalent to optimizing the following cost function
\begin{align}
    \mathcal{L} = \mathbb{E}_{t, \pmb{x}_0, \pmb{x}(t)}\{W\cdot||\pmb{\theta}(\pmb{x}(t)) - \pmb{\nabla}_{\pmb{x}}\ln(p_{t}(\pmb{x}))||^2\}
\end{align}
where $\pmb{\theta}(\pmb{x}(t))$ is the output of the neural network.
For interpolation-based diffusion,
\begin{align}
    p_{t}(\pmb{x})=p(\pmb{x}(t)|\pmb{x}_0, \pmb{y})=\mathcal{N}(\pmb{m}(\pmb{x}_0, \pmb{y}); G(t)\pmb{I})
\end{align}
where $p_{t}(\pmb{x})$ is the conditional probability density function of $\pmb{x}(t)$,  $\pmb{m}(\pmb{x}_0, \pmb{y})={\alpha}_t[\lambda _t\pmb{x}_0+(1-\lambda _t)\pmb{y}]$ is the mean of  $p_{t}(\pmb{x})$.
\begin{align}
    \pmb{\nabla}_{\pmb{x}}\ln(p_{t}(\pmb{x}))=\pmb{\nabla}_{\pmb{x}}[-\frac{||\pmb{x}(t) -\pmb{m}(\pmb{x}_0, \pmb{y})||^2}{2G^2(t)}]=-\frac{\pmb{z}}{G(t)}
\end{align}
here $\pmb{x}(t) -\pmb{m}(\pmb{x}_0, \pmb{y})=G(t)\pmb{z}$. 
Then we get the loss function
\begin{align}
    \mathcal{L} = \mathbb{E}\{W\cdot||\pmb{\theta}(\pmb{x}(t)) + \frac{\pmb{z}}{G(t)})||^2\}=\mathbb{E}||G(t)\pmb{\theta}(\pmb{x}(t)) + \pmb{z})||^2
    \label{loss}
\end{align}
We follow the settings in \cite{ song2020score, song2019generative, welker2022speech}, utilize the weighted loss, and set  $W=G^2(t)$ for better performance.

The training algorithm is shown in Alg. \ref{training}, where $\epsilon$ represents the minimum sample time, where the superscript $b$ in $[\cdot]^b$ denotes the $b$-th sample of a batch, batch size is $B$, $p(\pmb{x_0}, \pmb{y})$ denotes the joint probability density function of the clean and noisy pair, i.e., $\pmb{x}_0, \pmb{y}$.

\begin{table*}[th]
		
		\caption{Comparison of the VEIDM, VPIDM, and IDM.}
        \resizebox{17cm}{!}{%
		\label{compare}
		\centering
		\begin{tabular}{ ccc}
			\hline
			{ } & 
			{The state equations} &
			{The stochastic differential equations} \\
			\hline	
			VEIDM \cite{welker2022speech} & \begin{minipage}{11cm}{\begin{align}
			\pmb{x}(t) = \lambda _t\pmb{x}_0+(1-\lambda _t)\pmb{y} +\sqrt{\ln(\frac{\sigma_{\text{max}}}{\sigma_{\text{min}}}) \frac{\sigma_{\text{min}}^2((\sigma_{\text{max}}/\sigma_{\text{min}})^{2t}-e^{-2\lambda t})}{\lambda + \ln(\sigma_{\text{max}}/\sigma_{\text{min}})}} \pmb{z} \label{vet}
			\end{align}}  \end{minipage}  & \begin{minipage}{9cm}{\begin{align}
			    d\pmb{x}(t) = \lambda(\pmb{y}-\pmb{x}(t))dt +\sigma_{\text{min}}(\frac{\sigma_{\text{max}}}{\sigma_{\text{min}}})^t\sqrt{2\ln(\frac{\sigma_{\text{max}}}{\sigma_{\text{min}}})}d\pmb{w}
                \label{ved}
			\end{align}} \end{minipage} \\  		
			\hline	
   	VPIDM  & \begin{minipage}{11cm}{\begin{align}
			\pmb{x}(t) = {\alpha}_t[\lambda _t\pmb{x}_0+(1-\lambda _t)\pmb{y}] + \sqrt{1 - {\alpha}_t^2}\pmb{z} \label{vpt}
			\end{align}}  \end{minipage}  & \begin{minipage}{9cm}{\begin{align}
			   d\pmb{x}(t) = [ \pmb{x}(t)\ln^{'}{(\alpha}_t\lambda _t)-\pmb{y}{\alpha}_t\ln^{'}\lambda _t]dt + \pmb{\Sigma} _td\pmb{w}
                \label{vpd}
			\end{align}} \end{minipage} \\  		
			\hline	
   IDM  & \begin{minipage}{11cm}{\begin{align}
			  \pmb{x}(t) = {\alpha}_t[\lambda _t\pmb{x}_0+(1-\lambda _t)\pmb{y}] + G(t)\pmb{z} \label{dmt}
			\end{align}}  \end{minipage}  & \begin{minipage}{9cm}{\begin{align}
			   d\pmb{x}(t) = \pmb{x}(t)d\ln{(\alpha}_t\lambda _t)-\pmb{y}{\alpha}_td\ln\lambda _t + g(t)d\pmb{w}
                \label{dmd}
			\end{align}} \end{minipage} \\  		
			\hline	
		\end{tabular}%
  }
	\end{table*}

\subsection{The reverse process for sampling a clean}
From  \cite{song2020score, anderson1982reverse}, the reverse process is also a diffusion process which can be represented as

\begin{align}
    d\pmb{x}(t) = [f(\pmb{x}(t), \pmb{y}) - g(t)^2\pmb{\theta}(\pmb{x}(t))]dt + g(t)d\Bar{\pmb{w}}
\end{align}
where $f(\pmb{x}(t), \pmb{y})=\pmb{x}(t)\ln^{'}{(\alpha}_t\lambda _t)-\pmb{y}{\alpha}_t\ln^{'}\lambda _t$ for the VPIDM and $\Bar{\pmb{w}}$ is another stochastic process but has the same distribution with $\pmb{w}$. Typically, the continuous process is discretized  in the sampling stage.  $\Delta=\frac{1-\epsilon}{K}$ is set and let $\pmb{x}_k = \pmb{x}(\frac{k(1-\epsilon)}{K}+\epsilon)$ and $g_k = g(\frac{k(1-\epsilon)}{K}+\epsilon)\sqrt{\Delta}$. The sampling stage is elaborated in Alg. \ref{sampling}.
\begin{align}
    \pmb{x}_{k-1} =  \pmb{x}_{k} - [f(\pmb{x}_k, \pmb{y}) - g_k^2\pmb{\theta}(\pmb{x}_k)]\Delta + g_k{\pmb{z}}
    \label{resample}
\end{align}
where the subscript $k$ of $\pmb{x}_k$ and $g_k$ denote the discrete sampling time index,  $\pmb{x}_k$ and $g_k$ represent the discrete samplings of $\pmb{x}(t)$ and $g(t)$.


\subsection{Comparison with the VEIDM}
Furthermore, 
the proposed VPIDM and the VEIDM proposed in \cite{welker2022speech} can be concluded  as
\begin{align}
  d\pmb{x}(t) = \pmb{x}(t)d\ln{(\alpha}_t\lambda _t)-\pmb{y}{\alpha}_td\ln\lambda _t + g(t)d\pmb{w}
  \label{equation:13}
\end{align}

\begin{align}
  \pmb{x}(t) = {\alpha}_t[\lambda _t\pmb{x}_0+(1-\lambda _t)\pmb{y}] + G(t)\pmb{z}
  \label{equation:14}
\end{align}
The relation between $G$ and $g$ is constrained by Eq.\eqref{equation:12} which is referred to as the interpolation diffusion model (IDM). When $G^2(t)=1-\alpha^2_t$, the interpolation diffusion becomes a VP-based method. In the case of VE-based interpolation diffusion, $\alpha_t$ in Eq.\eqref{equation:13} and \eqref{equation:14} is constant $1$. Substitute $\alpha_t$ with constant $1$ and  solve the ordinary differential equation in Eq.\eqref{equation:12}, we get $G(t)$
\begin{align}
    \label{equation:15}
    G^2(t) = \lambda^2_tG(0)^2 + \lambda^2_t\int_0^t{g^2(\tau)}/{\lambda^2_\tau}d\tau
\end{align}
In \cite{welker2022speech}, The interpolation coefficient $\lambda _t$ is defined as $e^{-\lambda t}$. It can be verified that the VEIDM,  which use $G(t)$ from Eq.\eqref{vet}, and the $g(t)$ from Eq.\eqref{ved},  satisfies Eq.\eqref{equation:15} as a special case. Additionally, a comparison is made between the VEIDM, the proposed VPIDM, and IDM in Tab.\ref{compare}.

In the reverse process, the initial sample of the VEIDM in \cite{welker2022speech} is taken as $\pmb{y}+G(1)\pmb{z}$, rather than the ground truth $\lambda_1\pmb{x}_0+(1-\lambda_1)\pmb{y} + G(1)\pmb{z}$ because obtaining the clean is not possible  at this stage. However, this can result in damage to the enhanced speech. To quantify this effect, we define the initial error ($\textit{IE}$)
\begin{align}
    \textit{IE}_{\text{VEIDM}} &= [\pmb{y}+G(1)\pmb{z}] - [\lambda_1\pmb{x}_0+(1-\lambda_1)\pmb{y} + G(1)\pmb{z}] \\
    &= \lambda_1(\pmb{y}-\pmb{x}_0)
\end{align}
Whereas, the reverse of the proposed start from $\alpha_1\pmb{y}+G(1)\pmb{z}$, and the truth is $\alpha_1(\lambda_1\pmb{x}_0+(1-\lambda_1)\pmb{y}) + G(1)\pmb{z}$. The $\textit{IE}$ is 
\begin{align}
    \textit{IE}_{\text{VPIDM}} &= 
    \alpha_1\lambda_1(\pmb{y}-\pmb{x}_0)
\end{align}
When the same $\lambda_t$ is utilized in the Sgmse+ and VPIDM, $\textit{IE}_{\text{VPIDM}}\ll \textit{IE}_{\text{Sgmse+}}$, where $\alpha_1 \rightarrow 0$. That is, the VPIDM has a smaller $\textit{IE}$ than the VEIDM.

\section{Experiments}
\label{exp}
\subsection{Training settings}
We conduct our experiments on the publicly available benchmark, i.e., VoiceBank-DEMAND(VBD) \cite{valentini2016investigating} 
Metrics in \cite{le2019sdr, rix2001perceptual, 
hu2007evaluation}, 
i.e., SI-SDR, SI-SIR, SI-SAR, PESQ, CSIG, CBAK, COVL,
are adopted to 
compare the performance to other state-of-the-art methods.
$25$ speech clips from the test dataset are selected randomly as the validation dataset. We train the neural network for 120 epochs. The best checkpoint is saved when PESQ is in its optimal state during the validation phase.

We use the neural network proposed in \cite{richter2022speech} as our backbone model, which is originally introduced in \cite{song2020score} for the image generation task. 
We treat the complex spectrum as a real-valued tensor to circumvent complex-valued computation where the real and imaginary parts of the complex are represented as  two channels of the tensor. The tensor is scaled to ensure that its amplitude approximately falls within the range of $-1$ to $1$ before being fed into the neural network. We follow the scaling function described in \cite{welker2022speech} where given a complex-valued spectrum $\pmb{x}(t)=|\pmb{x}(t)|e^{\angle\pmb{x}(t)}$,
the scaled $[\pmb{x}(t)]^s=a|\pmb{x}(t)|^ce^{\angle\pmb{x}(t)}$, here $a, c$ are two hyper-parameters, $[\cdot ]^s$ means the operation of the scaling function.
We find that $a50^c\approx1, 0<c\leq1$ is of avail for the performance and  $c$ can not be set too small,  as it compresses the dynamic range of the signal drastically and  makes it more difficult to learn the structure of clean spectrum. In this paper, we empirically set $a=0.15$ and $c=0.5$,
 $\beta_{\text{min}}=0.1$,  $\beta_{\text{max}}=2$ and $\lambda_t = e^{-\lambda t}$, $\lambda=1.5$, $G(t)=\sqrt{1-\alpha_t^2}$.

\begin{figure*}[t]
  \centering
  \includegraphics[width=\linewidth]{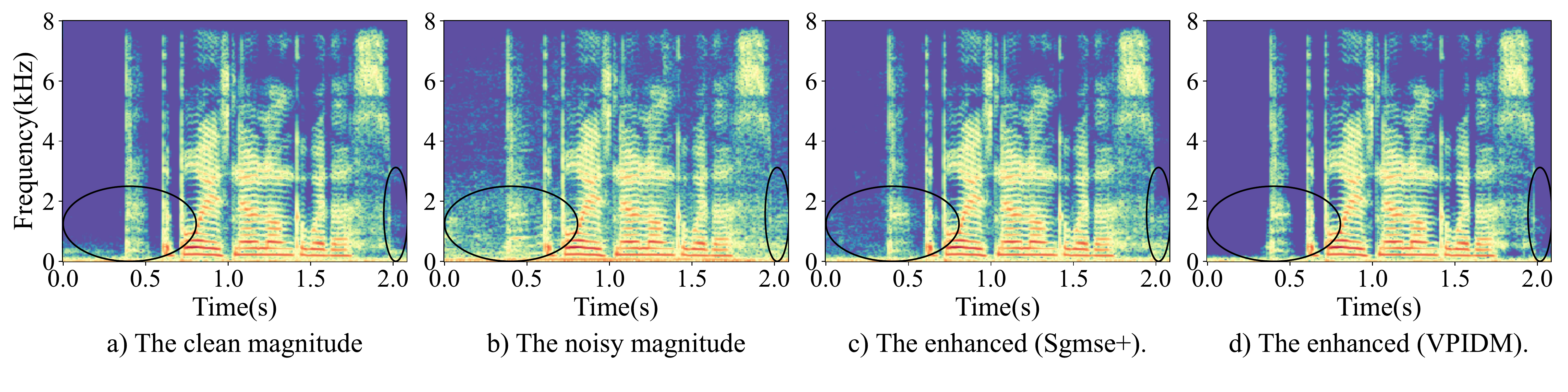}
  \caption{The log spectrogram of the STFT magnitudes. a) the clean $\pmb{x}_0$ spectrogram; b) the noisy $\pmb{y}$ spectrogram; c) 
 the spectrogram of the estimated clean from the Sgmse+ (2); d) the spectrogram of the estimated clean from the VPIDM.}
  \label{fig:mag}
\end{figure*}

%

\subsection{Results and analyses}
For the continuum diffusion model, we typically sample $t$ from the uniform distribution $\mathcal{U}(\epsilon, 1)$, $\epsilon$ denotes the time index of the first state after the clean. Ideally, we want to set $\epsilon$ as small as possible.
 However, when $\epsilon$ is too small, it can lead to difficulties in achieving convergence during training and may result in unstable fluctuations in the optimization process. 

	\begin{table}[th]
		
		\caption{Illustration of several metrics  over different settings for VPIDM on the VBD dataset.}
		\label{tab:ablations}
		\centering
		\resizebox{\columnwidth}{!}{%
		\begin{tabular}{ lcccc}
			\hline
			{Settings} & 
			{PESQ $\uparrow$} &
			{SISDR $\uparrow$} &
			{SISIR $\uparrow$} &
			{SISAR $\uparrow$} \\
			\hline
					
			$\epsilon= 1 \cdot 10^{-2}$ &  $3.01$   &${18.3}$ &$\mathbf{31.9}$ &$18.6$  \\
          
        $\epsilon= 3\cdot 10^{-2}$   & $3.02$    &$\mathbf{18.9}$ &$30.1$ &${19.4}$  \\	
		
			$\epsilon= 4\cdot 10^{-2}$   &$\mathbf{3.13}$    &$18.7$ &$28.6$ &$19.3$   \\			
			$\epsilon= 5\cdot10^{-2}$    & $2.86$ &$18.6$ &$27.6$ &$19.4$  \\
			$\epsilon= 6\cdot10^{-2}$   & $2.95$   &$18.7$ &$26.6$ &$\mathbf{19.7}$   \\
			$\epsilon= 7\cdot10^{-2}$   & $1.77$    &$16.1$ &$24.5$ &$16.9$    \\		
			
			\hline
			
		\end{tabular} %
  }
	\end{table}

\begin{figure}[t]
  \centering
  \includegraphics[width=\linewidth]{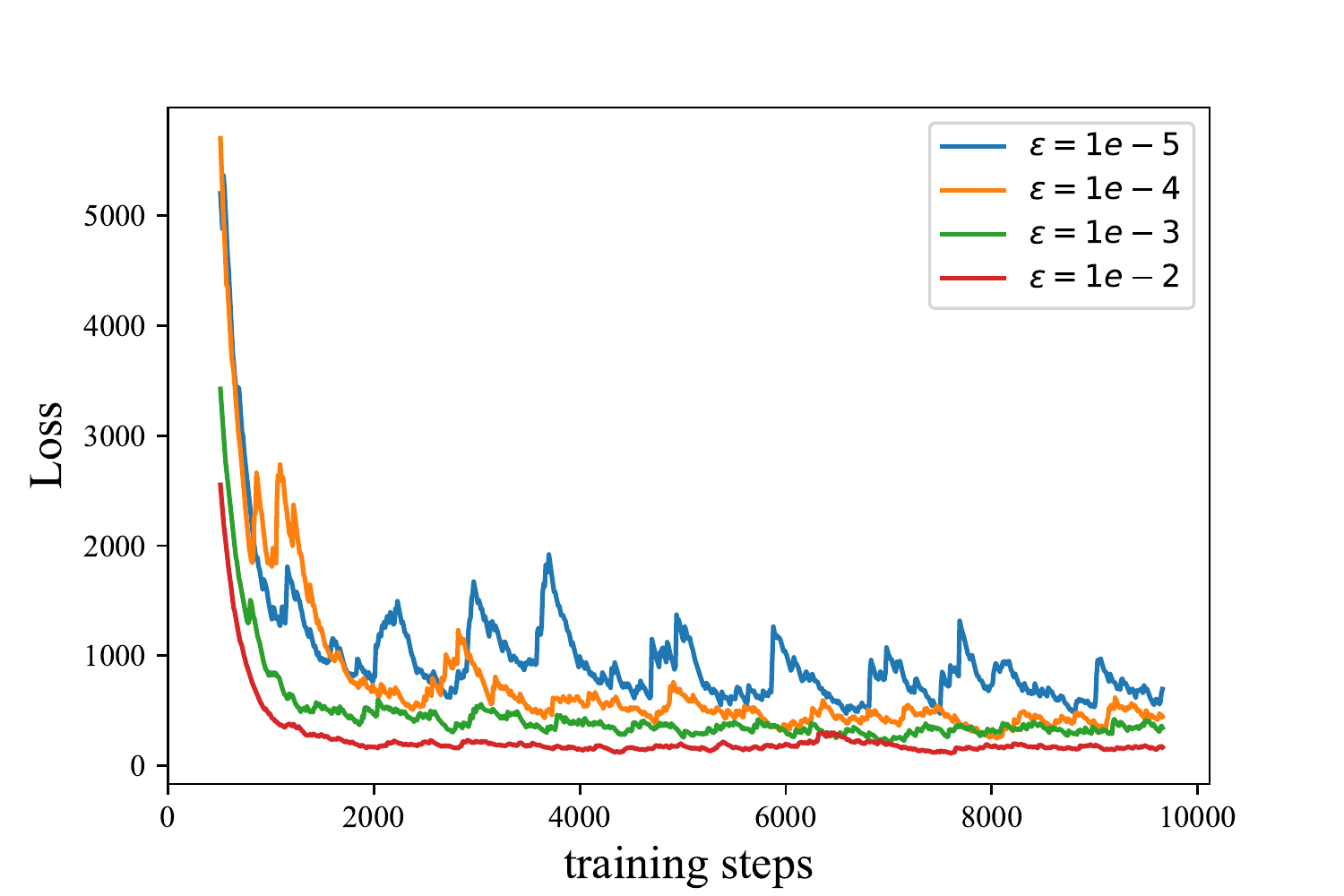}
  \caption{The training loss when $t\sim \mathcal{U}(\epsilon, 1)$.}
  \label{fig:training_loss}
\end{figure}

In Fig.\ref{fig:training_loss}, we vary the value of $\epsilon$ for the model and observe the training loss with training steps. The results show that when the $\epsilon$ is too small, the training loss exhibits lots of fluctuations and becomes difficult to converge. In our view, the reason is that, when $\epsilon$ is too small, the model is required to estimate the target in a wider range of SNR conditions. We conduct experiments to check the model's ability to predict the target in low SNR conditions using only one state ($t=\epsilon$) as the input for the model. However, the model can not predict the target well when $\epsilon \leq 10^{-2}$. That is,  when $t\rightarrow0$, $\alpha_t\rightarrow1$, $\lambda_t\rightarrow1$, then $\pmb{x}(t)\approx \pmb{x}_0+\sqrt{1-\alpha_t^2}\pmb{z}$. From the perspective of Gaussian noise, the $\text{SNR}\approx10\log_{10}(\frac{1-\alpha_t^2}{1})\approx 10\log_{10}(\beta_{\text{min}}t)$. Therefore, when $t=10^{-5}, 10^{-4}, 10^{-3}, 10^{-2}$, the model predicts the target at  about $-60$dB, $-50$dB, $-40$dB, $-30$dB SNR, respectively. Based on these results, we choose $\epsilon \geq 10^{-2}$. We treat the $\epsilon$ as the minimum resolution of the model, so we set the number of sample steps in the reverse process to $K\approx [\frac{1}{\epsilon}]$. As a result, for $\epsilon=[10^{-2},3\cdot10^{-2}, 4\cdot 10^{-2},5\cdot10^{-2},6\cdot10^{-2},10^{-1}]$, the number of sampling steps are $100, 30, 25, 20, 15, 10$, respectively.
According to Tab.\ref{tab:ablations},  the best PESQ is achieved when $\epsilon=4\cdot10^{-2}$, which corresponds to $25$ sample steps. The model demonstrates strong performance across various evaluation metrics when the number of sample steps exceeds $15$, but when the sample steps are less than $15$, the performance drops significantly.

\begin{table}[ht]
	
	\caption{The proposed method versus some SOTA
		methods with respect to different metrics.}
	\label{tab:overall}
	\centering
	\resizebox{\columnwidth}{!}{%
	\begin{tabular}{ lcccc}
		\hline
		{Model} & 
		
		{PESQ $\uparrow$} &
		{CSIG $\uparrow$} &
		{CBAK $\uparrow$} &
		{COVL $\uparrow$}  \\
		\hline	
           noisy   &$1.97$  &$3.35$    &$2.44$ &$2.64$  \\
		\hline
            PFP \cite{hsieh2020improving}    &$\mathbf{3.15}$   &$4.18$ &$\mathbf{3.60}$ &$3.67$   \\
            MetricGAN \cite{fu2019metricgan}   &$2.86$ &$3.99$    &$3.18$ &$3.42$     \\	
		MetricGAN$+$ \cite{fu21_interspeech}   &$3.15$   & $4.14$ &$3.16$ &$3.64$    \\
		CDiffuSE \cite{lu2022conditional}   & $2.52$  &$3.72$    &$2.91$ &$3.10$    \\	
		
		SRTNet \cite{srtnet} & $2.69$ & $4.12$   &$3.19$ &$3.39$    \\
			
		CDSE \cite{yen2022cold}   & $2.77$   & $3.91$ &$3.32$ &$3.33$   \\
            Sgmse+ (1) \cite{richter2022speech}  & $2.80$ & $4.10$    &$3.24$ &$3.44$   \\
            Sgmse+ (2) \cite{richter2022speech}  &$2.93$ &$4.12$    &$3.37$ &$3.51$   \\	
		
		\hline
		VPIDM & $3.13$ & $\mathbf{4.63}$    &$3.41$ &$\mathbf{3.94}$   \\
		\hline
		
	\end{tabular}%
 }
\end{table}


In Tab.\ref{tab:overall},  we compare our model to several SOTA methods. Sgmse+ (1) denotes the vanilla model without the corrector, while Sgmse+ (2) includes the corrector. The proposed model achieves the best CSIG, indicating the least speech distortion. This demonstrates that our generative model has learned the closest clean distribution to the ground truth that can preserve clean speech best.   Sgmse+ (1) in \cite{richter2022speech} is the VE-based interpolation diffusion with $30$ sampling steps. 
The proposed method with  fewer steps, i.e., $25$ steps  ($\epsilon=4\cdot 10^{-2}$),  achieves $0.3$ PESQ increment over Sgmse+ (1). To further improve the performance, Sgmse+ (2) in \cite{richter2022speech} implements a corrector which requires $60$ total steps. The proposed requires less than half steps of the Sgmse+ (2) and obtains a $0.2$ PESQ improvement. In fact, the scaling of the signal, i.e., $\alpha_t$,  on the right side of Eq.\eqref{equation:10} is a type of data augmentation that is beneficial for the model to learn the intrinsic structure of clean speech. 
However,  as shown in Fig.\ref{fig:mag},   Sgmse+ (2)  causes  the neural network to consider some noises in $\pmb{y}$ as the clean when starting from $\pmb{y}+g(1)\pmb{z}$ in the sampling stage. Therefore, some noises in the noisy speech are not effaced thoroughly. From the  two black ellipses in  Fig.\ref{fig:mag}, we can see that the Sgmse+ (2) has residual noise in $0-0.6$(s) and  $2-2.1$(s) time intervals.

\section{Conclusions}
\label{conclude}
In this paper, we present the VP-based interpolation diffusion model in a continuous time system and summarize the VE- and VP-based interpolation diffusion models into a more concise framework called the IDM. The VE- and VP-based interpolation diffusion models serve as examples of the IDM. While we only apply the VPIDM to the SE task to showcase our proposed method, it is worth noting that the approach is a general method that can be used for other tasks as well.



\section{Acknowledgements}
This work was supported in part by the National Natural Science Foundation of China under Grant 62171427. We also thank Midea Group Co., Ltd for funding this work

\bibliographystyle{IEEEtran}
\bibliography{template}

\end{document}